\documentclass[onecolumn]{svjour2}                   

\smartqed  

\usepackage{graphicx}

\journalname{J Stat Phys}

\begin{document}

\title{Charge and Current Sum Rules in Quantum Media Coupled to Radiation}

\titlerunning{Charge and Current Sum Rules in Quantum Media}

\author{Ladislav \v{S}amaj}

\institute{L. \v{S}amaj \at
               \\ Laboratoire de Physique Th\'eorique, Universit\'e
                  de Paris-Sud, 91405 Orsay Cedex, France (Unit\'e Mixte
                  de Recherche No 8627 - CNRS)
                \email{Ladislav.Samaj@th.u-psud.fr}  \\
                \emph{On leave from:} Institute of Physics,
                Slovak Academy of Sciences, Bratislava, Slovakia   
                \email{Ladislav.Samaj@savba.sk}  }

\date{Received:  / Accepted: }

\maketitle

\begin{abstract}
This paper concerns the equilibrium bulk charge and current density 
correlation functions in quantum media, conductors and dielectrics, 
fully coupled to the radiation (the retarded regime).
A sequence of static and time-dependent sum rules, which fix the values 
of certain moments of the charge and current density correlation functions,
is obtained by using Rytov's fluctuational electrodynamics.
A technique is developed to extract the classical and purely 
quantum-mechanical parts of these sum rules.
The sum rules are critically tested in the classical limit
and on the jellium model.
A comparison is made with microscopic approaches to systems of particles 
interacting through Coulomb forces only (the non-retarded regime).
In contrast with microscopic results, the current-current density correlation 
function is found to be integrable in space, 
in both classical and quantum regimes.


\end{abstract}

\renewcommand{\theequation}{1.\arabic{equation}}
\setcounter{equation}{0}

\section{Introduction} \label{Sect.1}
This paper is about the equilibrium statistical mechanics of infinite (bulk)
media, conductors and dielectrics, fully coupled to the radiated 
electromagnetic (EM) field.
The crucial problem of statistical analysis of such models is to determine 
how fluctuations of various statistical quantities, like charge and current 
densities, scalar and vector potentials, electric and magnetic fields, etc., 
around their mean values are correlated in space and time.
A special attention is devoted to the behavior of correlation functions
at asymptotically large distances and to the so-called sum rules, which fix 
the values of certain moments of the correlation functions.  

The studied models are composed of spinless charged particles which
are quantum, but non-relativistic (i.e., they obey the Schr\"odinger
equation and not the Dirac particle-antiparticle formalism);
according to the correspondence principle, a quantum system admits
the classical description in the high-temperature limit.
On the other hand, the interaction of charged particles via the radiated
EM field can be considered either non-relativistic (non-retarded) 
or relativistic (retarded).
In the non-retarded regime, the speed of light $c$ is taken to be infinitely 
large, $c=\infty$, ignoring in this way magnetic forces acting on 
the particles; in other words, charges interact pairwisely only via 
instantaneous Coulomb potentials.
In the retarded regime, $c$ is assumed finite and the particles are fully 
coupled to both electric and magnetic (transverse) parts of the radiated field.
The ultimate aim is to describe quantum particles in the retarded regime,
but due to serious technical difficulties one usually starts with the simpler
non-retarded regime. 

Two complementary approaches exist in the theoretical study of charged systems.
The microscopic one is based on the explicit solution of models defined by 
their microscopic Hamiltonians.
The macroscopic one is based on the assumption of validity of macroscopic 
electrodynamics.
Microscopic description is complicated and laborious, but if some explicit 
results are available their value is considerable and, e.g., they can reveal 
a restricted applicability of macroscopic methods.

Microscopic approaches are usually restricted to the non-retarded regime. 
A particular interest was devoted in the past to classical and quantum models 
of conductors. 
Classical particles interacting by the instantaneous Coulomb potential
have been treated in the pioneering work by Debye and H\"uckel \cite{Debye23}.
An exponential shielding of the Coulomb potential, typical for the classical 
regime, was observed.
One of the first classical analyses of a system of charged particles fully 
coupled to the radiation, based on the linearized Vlasov equation and 
Maxwell's equations, was performed by Felderhof \cite{Felderhof64,Felderhof65},
see also the monograph \cite{VanKampen67}.  
The existence of general sum rules for certain moments of the classical charge 
density correlation functions was established by Stillinger and Lovett 
\cite{Stillinger68}.
The quantum sum rules, static and time-dependent, were the subject of 
numerous studies \cite{Pines66,Martin85,Martin86,Martin88,John93}.
A treatment of quantum Coulomb fluids, based on the use of the Feynman-Kac 
path integral representation of the thermal Gibbs weight \cite{Alastuey89}, 
indicates that the charge-charge density correlation functions exhibit 
a long-range $1/r^6$ decay (see review \cite{Brydges99}).
A microscopic analysis of conductor systems when transverse EM interactions 
are added to the Coulomb ones, was done recently 
\cite{Boustani06,Buenzli07}; a conceptual problem related to the use of 
classical radiation was pointed out in \cite{Jancovici06}.
Microscopic models of dielectrics are less developed; we mention
a fluid model of quantized polarizable particles \cite{Hoye81,Thompson82}
with the static dipole-dipole interaction.

Macroscopic phenomenological approaches usually allow us to predict, with less 
effort, basic features of relatively complicated complex physical systems
in the retarded regime.
A macroscopic theory of equilibrium thermal fluctuations of the EM
field in conductors and dielectrics was proposed by Rytov 
\cite{Rytov53,Rytov58,Levin67}, see also the Landau and Lifshitz  
Course of Theoretical Physics \cite{LP} and the book \cite{Sitenko95}.
For bulk systems, an interesting phenomenon occurs in two-point correlation
functions of microscopic quantities like the induced electric potential or 
field: the presence of transverse interactions beyond Coulomb cancel out
some (spatially nonintegrable) long-ranged terms observed in 
the non-retarded regime \cite{Jancovici06,LP}. 
Main applications of Rytov's theory were related to inhomogeneous situations.
In the famous Lifshitz paper \cite{Lifshitz56}, Rytov's theory was used 
in the context of the Casimir attraction between parallel dielectric plates.
Another application appeared in the field of thermally excited surface
EM waves \cite{Joulain05}.
Recently, the theory was used in the study of retardation effects
on the long-range surface charge density correlation function between
two distinct media \cite{Samaj08,Jancovici09}. 

Macroscopic approaches, being essentially of mean-field type,
are expected to provide reliable results for the leading terms
in the asymptotic long-wavelength behavior of correlations.
Those results can be recast as sum rules for moments of correlations 
integrated in real space, or as asymptotic large-distance behaviors.
As concerns these asymptotic behaviors, macroscopic mean-field teories
fail when fluctuations become crucial.
The failure of mean-field predictions is well illustrated by the
existence of $1/r^6$ tails in static charge-charge correlations
of quantum Coulomb systems, as explained in \cite{Alastuey99}; see also 
the exact calculation of the tail for hydrogen plasma at low densities 
\cite{Cornu97}.
Another implicit assumption for the validity of macroscopic approaches is
that correlations decay sufficiently fast.
It is then legitimate to perform local averages for EM fields slowly varying 
in space, which  reflects itself in the introduction of the frequency
dependent dielectric function $\epsilon(\omega)$ in Maxwell's equations.

The present paper concerns sum rules for the bulk charge and current density 
correlation functions in quantum media, conductors and dielectrics, 
fully coupled to the radiation.
Although this subject (in the non-retarded version) was in the center of 
interest of microscopic theories in the past, it remained practically 
unheeded by macroscopic approaches.
Here, a sequence of static or time-dependent sum rules, known or new, 
is obtained by Rytov's fluctuational electrodynamics.
A technique is developed to extract the classical and purely 
quantum-mechanical parts of these sum rules.
The sum rules are critically tested on a jellium model of conductors.
A comparison is made with microscopic approaches to systems of particles 
interacting through Coulomb forces only; in contrast to these microscopic 
approaches, the current-current density correlation function is found 
to be integrable in space, in both classical and quantum regimes.

The paper is organized as follows.
In Sect. 2, we review briefly Rytov's theory of EM field fluctuations and
present basic expressions for the charge and current densities.
The sum rules for the charge-charge, charge-current and current-current 
density correlation functions are derived and discussed in Sects. 3, 4 and 5,
respectively.
Sect. 6 deals with problematic points of both the Rytov fluctuational theory
and microscopic approaches.
Sect. 7 is devoted to an interesting comparison of the present results to 
those of Felderhof \cite{Felderhof64,Felderhof65}, in the classical limit.
Sect. 8 is the Conclusion.

\renewcommand{\theequation}{2.\arabic{equation}}
\setcounter{equation}{0}

\section{Macroscopic fluctuational formalism} \label{Sect.2}
We consider the $(3+1)$-dimensional space, defined by Euclidean vectors 
${\bf r}=(x_1,x_2,x_3)$ and time $t$.
The two-point functions considered in this paper will be translationally 
invariant in both Euclidean space and time, so we can use the spectral 
(Fourier) representation
\begin{equation} \label{2.1}
f(t,{\bf r};t',{\bf r}') \equiv f(t-t',{\bf r}-{\bf r}')
= \int \frac{{\rm d}\omega}{2\pi}
\int \frac{{\rm d}^3 q}{(2\pi)^3} {\rm e}^{-{\rm i}\omega(t-t')
+ {\rm i}{\bf q}\cdot({\bf r}-{\bf r}')} f(\omega,{\bf q}) ,
\end{equation}
where $\omega$ denotes the frequency and ${\bf q}=(q_1,q_2,q_3)$
the wave vector.
The physical system of interest is a medium (conductor or dielectric) and 
an EM field present in it, which are in thermal equilibrium.

The medium is composed of moving charged particles which are assumed
to be non-relativistic.
In the long-wavelength scale much larger than the mean interparticle distance, 
the macroscopic characteristic of the medium is the frequency dependent
dielectric function $\epsilon(\omega)$.
We shall assume that the matter has no magnetic structure, i.e. it is not
magnetoactive, and permeability $\mu=1$.
The medium is coupled to the EM field generated by the particles.
The {\em classical} EM field is described by the scalar potential 
$\phi(t,{\bf r})$ and the vector potential ${\bf A}(t,{\bf r})$
with component $A_j(t,{\bf r})$ $(j=1,2,3)$.
In the considered Weyl gauge $\phi=0$, the microscopic electric and
magnetic fields are given by
\begin{eqnarray}
{\bf E} & = & - \frac{1}{c} \frac{\partial {\bf A}}{\partial t} , 
 \label{2.2} \\
{\bf B} & = & {\rm curl}\, {\bf A} , \qquad
B_j = \sum_{k,l} e_{jkl} \frac{\partial}{\partial x_k} A_l \label{2.3}
\end{eqnarray}
with $c$ being the velocity of light and $e_{jkl}$ being the unit 
antisymmetric pseudotensor.
The elementary excitations of the {\em quantized} EM field are described
by the photon operators of the vector potential components $\hat{A}_j$  
which are self-conjugate Bose operators.
The retarded photon Green function tensor ${\bf D}$ is defined by
\begin{equation} \label{2.4}
{\rm i} D_{jk}(t-t';{\bf r},{\bf r}') = \left\{
\begin{array}{lr} 
\langle \hat{A}_j(t,{\bf r}) \hat{A}_k(t',{\bf r}') -
\hat{A}_k(t',{\bf r}') \hat{A}_j(t,{\bf r}) \rangle , & t\ge t' ,\cr
& \cr 0 , & t<t' .
\end{array} \right. 
\end{equation}
Here, $\hat{A}_j(t,{\bf r})$ denotes the vector-potential operator
in the Heisenberg picture and the angular brackets represent
the equilibrium averaging at temperature $T$.
For non-magnetoactive media, the Green function tensor possesses
the symmetry
\begin{equation} \label{2.5}
D_{jk}(t-t';{\bf r},{\bf r}') = D_{kj}(t-t';{\bf r}',{\bf r}) .
\end{equation}

The EM fields are random variables which fluctuate around their
mean values governed by Maxwell's equations. 
To describe fluctuations of the EM fields, Rytov 
\cite{Rytov53,Rytov58,Levin67,LP} studied the effect of 
a {\em weak} classical current ${\bf j}(t,{\bf r})$ due to 
the random particle motion in the medium.
This current acts as an ``external force'' on the vector-potential 
operator in the interaction Hamiltonian
\begin{equation} \label{2.6}
{\cal H}_{\rm int}(t) = - \frac{1}{c} \int {\rm d}^3 r\,
{\bf j}(t,{\bf r}) \cdot \hat{\bf A} .
\end{equation}
The mean values of the components of the vector-potential operator
can be thus expressed in terms of the Green function (\ref{2.4})
by using Kubo's linear response in current,
\begin{equation} \label{2.7}
\frac{\bar{A}_j(t,{\bf r})}{c} = - \frac{1}{\hbar c^2} 
\int {\rm d}^3 r'\int {\rm d}t'\, \sum_k D_{jk}(t-t';{\bf r},{\bf r}') 
j_k(t',{\bf r}') .
\end{equation} 
The mean value $\bar{\bf A}$ satisfies macroscopic Maxwell's equations 
due to the classical current ${\bf j}$.
This fact automatically implies a set of differential equations of dyadic type
fulfilled by the Green function tensor.
In particular, in the frequency Fourier space, we have
\begin{equation} \label{2.8}
\sum_l \left[ \frac{\partial^2}{\partial x_j \partial x_l}
- \delta_{jl} \Delta - \delta_{jl} \frac{\omega^2}{c^2} 
\epsilon(\omega) \right] D_{lk}(\omega;{\bf r},{\bf r}') 
= - 4 \pi \hbar \delta_{jk} \delta({\bf r}-{\bf r}') .
\end{equation} 
This set of equations has to be supplemented by the obvious boundary 
condition of regularity $D_{lk}(\omega;{\bf r},{\bf r}')\to 0$
for the distance $\vert {\bf r}-{\bf r}'\vert$ going to infinity.
In the case of a spatially homogeneous infinite medium, 
$D_{jk}(t-t';{\bf r},{\bf r}') \equiv D_{jk}(t-t';{\bf r}-{\bf r}')$, 
under the Fourier transform with respect to the difference ${\bf r}-{\bf r}'$ 
Eq. (\ref{2.8}) reduces to a set of algebraic equations, whose solution reads
\begin{equation} \label{2.9}
D_{jk}(\omega,{\bf q}) = \frac{4\pi\hbar c^2}{\omega^2\epsilon(\omega)}
\frac{q_j q_k}{q^2} + \frac{4\pi\hbar c^2}{\omega^2\epsilon(\omega)-(cq)^2}
\left( \delta_{jk} - \frac{q_j q_k}{q^2} \right) .
\end{equation}
Here, $q=\vert {\bf q}\vert$.
The first/second term on the right-hand-side (r.h.s.) of this expression 
represents the longitudinal/transverse component of the Green function.  
Formula (\ref{2.9}) does reduce to its well-known form in the vacuum
when one sets $\epsilon(\omega)=1$.

Eq. (\ref{2.7}) defines $-D_{jk}(\omega;{\bf r},{\bf r}')/(\hbar c^2)$ 
as the tensor of susceptibilities corresponding to the random variable
$\hat{\bf A}/c$.
The fluctuation-dissipation theorem tells us that the fluctuations
of random variables are expressible in terms of the corresponding
generalized susceptibilities. 
For the present symmetry (\ref{2.5}), the theorem implies 
\begin{equation} \label{2.10}
\langle A_j A_k \rangle_{\omega,{\bf q}}^{s} = - 
\coth\left( \frac{\beta\hbar\omega}{2} \right) 
{\rm Im}\, D_{jk}(\omega,{\bf q}) ,
\end{equation}
where $\beta=1/(k_{\rm B}T)$ is the inverse temperature and 
$\langle A_j A_k \rangle_{\omega,{\bf q}}^{s}$ is the Fourier
transform of the symmetrized correlation function
\begin{equation} \label{2.11}
\langle \hat{A}_j(t,{\bf r}) \hat{A}_k(t',{\bf r}') \rangle^{s} \equiv
\frac{1}{2} \langle \hat{A}_j(t,{\bf r}) \hat{A}_k(t',{\bf r}')
+ \hat{A}_k(t',{\bf r}') \hat{A}_j(t,{\bf r}) \rangle .
\end{equation}
We note that the symmetrization (\ref{2.11}) is needless in the classical case.

The fluctuation formula (\ref{2.10}) enables us to calculate the correlations 
of arbitrary statistical quantities provided that they are expressible 
in terms of the components of the vector-field potential, within the
classical EM field and in the gauge defined by 
Eqs. (\ref{2.2}) and (\ref{2.3}).
Let a scalar quantity $u$ be expressible as $u(t,{\bf r}) = 
{\bf U} \cdot {\bf A}(t,{\bf r}) \equiv \sum_j U_j A_j(t,{\bf r})$,
where $U_j$ are operators acting on $t$ and ${\bf r}$ variables.
Within the spectral representation with a single frequency and wave vector,
$f(t,{\bf r}) = {\rm e}^{-{\rm i}\omega t + {\rm i}{\bf q}\cdot{\bf r}}
f(\omega,{\bf q})$, this relation takes an algebraic form
$u(\omega,{\bf q})=\sum_j U_j(\omega,{\bf q}) A_j(\omega,{\bf q})$.
It follows from the definition (\ref{2.1}) that the Fourier transform of 
the symmetrized correlation function of statistical quantities 
$u$ and $v$, $\langle u(t,{\bf r}) v(t',{\bf r}') \rangle^s$, 
is then determined by
\begin{eqnarray}
\langle u v \rangle^s_{\omega,{\bf q}} & = &
\langle u(\omega,{\bf q}) v(-\omega,-{\bf q}) \rangle^s \nonumber \\
& = & \sum_{j,k} U_j(\omega,{\bf q}) V_k(-\omega,-{\bf q}) 
\langle A_j(\omega,{\bf q}) A_k(-\omega,-{\bf q}) \rangle^s , \label{2.12}
\end{eqnarray}
where $\langle A_j(\omega,{\bf q}) A_k(-\omega,-{\bf q}) \rangle^s
\equiv \langle A_j A_k \rangle^s_{\omega,{\bf q}}$ given by (\ref{2.10}).

In this paper, we shall concentrate on the volume charge density $\rho$ and 
the electric current density ${\bf j}$.
The charge density is expressible as
\begin{equation} \label{2.13}
\rho(t,{\bf r}) = \frac{1}{4\pi} {\rm div}\, {\bf E}
= - \frac{1}{4\pi c} \sum_j \frac{\partial}{\partial x_j}
\frac{\partial}{\partial t} A_j(t,{\bf r})
\end{equation}
or, in the spectral representation, as
\begin{equation} \label{2.14}
\rho(\omega,{\bf q}) = - \frac{\omega}{4\pi c} \sum_j q_j A_j(\omega,{\bf q}) .
\end{equation}
The electric current density is expressible in terms of the electric and
magnetic fields as follows
\begin{equation} \label{2.15}
{\bf j}(t,{\bf r}) = \frac{c}{4\pi} {\rm curl}\, {\bf B}(t,{\bf r})
- \frac{1}{4\pi} \frac{\partial}{\partial t} {\bf E}(t,{\bf r}) .
\end{equation}
Substituting here the gauge relations (\ref{2.2}) and (\ref{2.3}), and using 
the summation formula for the elements of the unit antisymmetric pseudotensor
\begin{equation} \label{2.16}
\sum_{l=1}^3 e_{jkl} e_{lmn} = \delta_{jm} \delta_{kn} - \delta_{jn} 
\delta_{km} 
\end{equation}
we obtain
\begin{equation} \label{2.17}
j_k(t,{\bf r}) = \frac{c}{4\pi} \left[ 
\sum_l \frac{\partial^2}{\partial x_l \partial x_k} A_l(t,{\bf r}) -
\sum_l \frac{\partial^2}{\partial x_l^2} A_k(t,{\bf r}) \right]
+ \frac{1}{4\pi c} \frac{\partial^2}{\partial t^2} A_k(t,{\bf r}) 
\end{equation}
or, in the spectral representation,
\begin{equation} \label{2.18}
j_k(\omega,{\bf q}) = \frac{c}{4\pi} \left[ - q_k \sum_l q_l 
A_l(\omega,{\bf q}) +
\left( q^2 - \frac{\omega^2}{c^2} \right) A_k(\omega,{\bf q}) \right] .
\end{equation}

It is a simple task to verify that both versions of the continuity equation
\begin{equation} \label{2.19}
\frac{\partial}{\partial t} \rho(t,{\bf r}) + 
{\rm div}\, {\bf j}(t,{\bf r}) = 0 , \qquad
-\omega \rho(\omega,{\bf q}) + \sum_k q_k j_k(\omega,{\bf q}) = 0
\end{equation}
are satisfied.

Henceforth, in order to simplify the notation we set $t'=0$ and
${\bf r}'={\bf 0}$, i.e. the time and position differences
between the two points in the correlation functions will be 
$t$ and ${\bf r}$, respectively.

\renewcommand{\theequation}{3.\arabic{equation}}
\setcounter{equation}{0}

\section{Charge-charge density correlations} \label{Sect.3}
We start with the charge-charge density correlation function.
Using the spectral representation (\ref{2.14}) it is easy to show that
\begin{equation} \label{3.1}
\beta \langle \rho \rho \rangle_{\omega,{\bf q}}^s = 
\beta \left( \frac{\omega}{4\pi c} \right)^2 \sum_{j,k} q_j q_k 
\langle A_j A_k \rangle_{\omega,{\bf q}}^s
= - q^2 \frac{g(\omega)}{2\pi\omega} 
{\rm Im}\, \frac{1}{\epsilon(\omega)} ,
\end{equation}
where
\begin{equation} \label{3.2}
g(\omega) \equiv \frac{\beta\hbar\omega}{2} 
\coth\left( \frac{\beta\hbar\omega}{2} \right) .
\end{equation}
Note that $g(\omega)\ge 1$; the equality $g(\omega)=1$ takes place
in the limit $\beta\hbar\omega\to 0$. 
The fact that there is only one term of the order $q^2$ in 
the Fourier representation (\ref{3.1}) is related to the applicability of 
the fluctuational theory in the long-wavelength limit 
$q*({\rm characteristic\ length})\to 0$ 
(characteristic length is usually the mean interparticle distance);
an exact formula for $\beta \langle \rho \rho \rangle_{\omega,{\bf q}}^s$,   
valid for any value of ${\bf q}$, would contain also higher powers of $q^2$.

The inverse Fourier transform of (\ref{3.1}) leads to 
the zeroth-moment (neutrality) condition
\begin{equation} \label{3.3}
\int {\rm d}^3 r\, \beta 
\langle \hat{\rho}(t,{\bf r}) \hat{\rho}(0,{\bf 0}) \rangle^s = 0
\end{equation}
and the second-moment condition
\begin{equation} \label{3.4}
\frac{1}{3} \int {\rm d}^3 r\, r^2 \beta 
\langle \hat{\rho}(t,{\bf r}) \hat{\rho}(0,{\bf 0}) \rangle^s
= \int_{-\infty}^{\infty} {\rm d}\omega\, 
\exp(-{\rm i}\omega t) \frac{g(\omega)}{2\pi^2\omega} 
{\rm Im}\, \frac{1}{\epsilon(\omega)} .
\end{equation}
These sum rules do not depend on $c$ and so they are the same
in both nonretarded and retarded regimes.
Note that the sum rule (\ref{3.4}) is not universal in the sense
that its r.h.s. is a complicated function of the temperature and
the medium characteristics. 

\subsection{Static case}
\begin{figure} \label{Fig.1}      
\includegraphics[width=0.60\textwidth,clip]{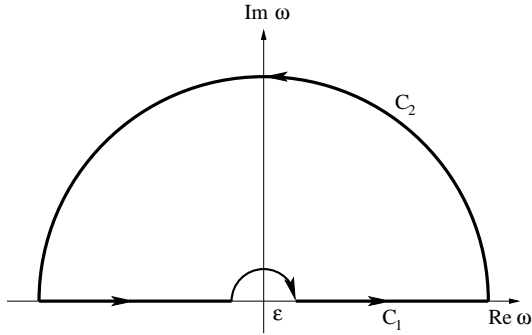}
\caption{The contour in the complex frequency plane.}
\end{figure}

\noindent In the static $t=0$ case, using the notation 
$\hat{\rho}(0,{\bf r}) \equiv \hat{\rho}({\bf r})$,
the general result (\ref{3.4}) can be
rewritten in the form
\begin{equation} \label{3.5}
\frac{1}{3} \int {\rm d}^3 r\, r^2 \beta 
\langle \hat{\rho}({\bf r}) \hat{\rho}({\bf 0}) \rangle^s
= \int_{-\infty}^{\infty} 
\frac{{\rm d}\omega}{\omega} {\rm Im}\, f(\omega) , \quad
f(\omega) = \frac{g(\omega)}{2\pi^2} 
\left[ \frac{1}{\epsilon(\omega)} - 1 \right] ; 
\end{equation}
note that the subtraction of 1 from $1/\epsilon(\omega)$ has no effect
due to the operation ${\rm Im}$.

The integral on the r.h.s. of Eq. (\ref{3.5}) can be formally evaluated by 
using integration techniques and the general analytic properties of dielectric 
functions in the complex frequency upper half-plane, in close analogy with 
the treatment of the surface charge density correlation function in 
Sect. IV of Ref. \cite{Jancovici09}.
We shall not repeat the derivation procedure and only write down the final
result. 
Although our $f$-function defined in (\ref{3.5}) differs from that 
considered in \cite{Jancovici09}, it possesses the necessary symmetry
$f^*(\omega) = f(-\omega)$ for real $\omega$ and goes to zero in 
the asymptotic limit $\vert\omega\vert\to\infty$ due to the asymptotic
relation \cite{Jackson,LL}
\begin{equation} \label{3.6}
\epsilon(\omega) \mathop{\sim}_{\vert\omega\vert\to\infty}
= 1 - \frac{\omega_p^2}{\omega^2} , \qquad 
\omega_p^2 = \sum_{\sigma} \frac{4\pi n_{\sigma} e_{\sigma}^2}{m_{\sigma}} .
\end{equation}
Here, $\omega_p$ is the plasma frequency of the medium composed of
species $\sigma$ (electrons and ions) with the number density $n_{\sigma}$,
charge $e_{\sigma}$ and mass $m_{\sigma}$;
since the masses of ions is much bigger than the mass of electron,
it is usually sufficient to approximate the plasma frequency of the medium
by $\omega_p^2 = 4\pi n e^2/m$ with $n$ being the number of all electrons 
in all atoms in unit volume, $e$ and $m$ are the electron charge and mass, 
respectively.
Let $C$ be a closed contour in the $\omega$ upper half-plane defined as
$C=C_1\cup C_2$, where $C_1$ is the path following the real axis, except
it goes around the origin $\omega=0$ in a small semicircle whose radius
$\epsilon$ tends to zero and $C_2$ is a semicircle at infinity (see Fig. 1).
If we denote by $\{ \omega_j\}$ all poles of the function $f(\omega)$
in the region bounded by the contour $C$, the procedure described in
Ref. \cite{Jancovici09} implies
\begin{equation} \label{3.7}
\int_{-\infty}^{\infty} 
\frac{{\rm d}\omega}{\omega} {\rm Im}\, f(\omega) = \pi f(0)
+2 \pi \sum_j \frac{{\rm Res}(f,\omega_j)}{\omega_j} , 
\end{equation}
where ${\rm Res}$ means the residue.
The dielectric function $\epsilon(\omega)$ has no zeros in the upper 
half-plane, so the only source of the poles of $f(\omega)$ inside $C$ is 
the function $g(\omega)$ given by (\ref{3.2}). 
Since $g(\omega)$ can be expanded in $\omega$ as follows \cite{Gradshteyn}
\begin{equation} \label{3.8}
g(\omega) \equiv \frac{\beta\hbar\omega}{2} 
\coth\left( \frac{\beta\hbar\omega}{2} \right) =
1 + \sum_{j=1}^{\infty} \frac{2\omega^2}{\omega^2+\xi_j^2} ,
\qquad \xi_j = \frac{2\pi}{\beta\hbar} j ,
\end{equation}
it exhibits in the upper half-plane an infinite sequence of simple poles
at the (imaginary) Matsubara frequencies
\begin{equation} \label{3.9}
\omega_j = {\rm i}\xi_j , \qquad {\rm Res}(g,\omega_j) = \omega_j
\quad (j=1,2,\ldots).
\end{equation}
We thus conclude that the second moment of the static charge-charge density 
correlation function (\ref{3.5}) is expressible as 
\begin{equation} \label{3.10}
\frac{1}{3} \int {\rm d}^3 r\, r^2 \beta 
\langle \hat{\rho}({\bf r}) \hat{\rho}({\bf 0}) \rangle^s = 
\frac{1}{2\pi} \left[ \frac{1}{\epsilon(0)}-1 \right] + \frac{1}{\pi}
\sum_{j=1}^{\infty} \left[ \frac{1}{\epsilon({\rm i}\xi_j)}-1 \right] . 
\end{equation}
In the $\omega$ upper half-plane, the dielectric function $\epsilon(\omega)$
takes real values only just on the imaginary axis and this fact ensures 
that the second moment (\ref{3.10}) is real.
Moreover, $\epsilon(\omega)$ decreases on the imaginary axis monotonically
from $\epsilon_0>1$ (for dielectrics) or from $\infty$ (for conductors) at 
$\omega={\rm i}0$ to 1 at $\omega={\rm i}\infty$; the convergence of the sum 
on the r.h.s. of (\ref{3.10}) is ensured by the asymptotic behavior (\ref{3.6}).

The representation (\ref{3.10}) is very useful in the high-temperature
(classical) limit $\beta\hbar\omega_p\to 0$ when each of the frequencies 
$\{ \xi_j \}_{j=1}^{\infty}$ is much larger than $\omega_p$, the corresponding 
terms in the sum on the r.h.s. of Eq. (\ref{3.10}) vanish and therefore
\begin{equation} \label{3.11}
\frac{1}{3} \int {\rm d}^3 r\, r^2 \beta 
\langle \rho({\bf r}) \rho({\bf 0}) \rangle_{\rm cl} = 
\frac{1}{2\pi} \left[ \frac{1}{\epsilon(0)}-1 \right] .
\end{equation}
In the case of conductors, setting $\epsilon(0)\to {\rm i}\infty$
implies the universal Stillinger-Lovett sum rule \cite{Stillinger68}.
In the case of a dielectric of static dielectric constant
$\epsilon(0) = \epsilon_0$, the above nonuniversal result has already been 
obtained by Chandler \cite{Chandler77}; see also the article \cite{Alastuey00}.
It should be noted that, at finite temperature, a quantum system always 
contains a finite fraction of ionized charges and thus, in a strict sense,
it is conductor.
In the case of the so-called dielectrics, the corresponding screening length
is huge, much larger than the size of considered macroscopic samples, and
therefore one does observe dielectric properties at experimental
length scales.
This point is well illustrated by exact results for atomic hydrogen
\cite{Ballenegger03}. 

The formula (\ref{3.11}) can be rederived for dielectrics in an alternative 
way by extending screening and linear response ideas in classical Coulomb 
fluids \cite{Jancovici95,Pines66} to dielectrics as follows \cite{Janco}.
If we put an infinitesimal charge $\delta e$ at position ${\bf r}$,
the dielectric - test charge interaction Hamiltonian reads 
$H_{\rm int} = \delta e \phi({\bf r})$, where $\phi({\bf r})$ is 
the potential at point ${\bf r}$ created purely by the dielectric.
By linear response theory, the average of this potential at some
point ${\bf r}'$ is changed by the presence of $\delta e$ by
\begin{equation} \label{3.12}
\langle \phi({\bf r}') \rangle_{\delta e}
= - \beta \langle H_{\rm int} \phi({\bf r}') \rangle 
= - \beta \delta e \langle \phi({\bf r}) \phi({\bf r}') \rangle ,
\end{equation}
where we suppose that $\langle \phi({\bf r}) \rangle = 0$.
The basic physical assumption about screening in dielectric media is that 
only a fraction $(1-1/\epsilon_0)$ of the external charge is screened. 
Consequently, $\delta e$ surrounds itself with a polarization cloud of
microscopic size carrying charge $-(1-1/\epsilon_0)\delta e$.
For $\vert {\bf r}-{\bf r}' \vert$ much larger than any microscopic scale,
the l.h.s. of (\ref{3.12}) is thus the potential at ${\bf r}'$ due 
to a pointlike charge $(1/\epsilon_0-1)\delta e$ at ${\bf r}$, i.e.
\begin{equation} \label{3.13}
\left( \frac{1}{\epsilon_0} - 1 \right)
\frac{\delta e}{\vert {\bf r}-{\bf r}' \vert}
= - \beta \delta e \langle \phi({\bf r}) \phi({\bf r}') \rangle .
\end{equation}
The charge density $\rho({\bf r})$ is related to the potential
$\phi({\bf r})$ via the Poisson equation $\Delta\phi = - 4\pi\rho$.
Taking first the Laplacian on ${\bf r}$ and then on ${\bf r}'$ in both
sides of Eq. (\ref{3.13}), we obtain
\begin{equation} \label{3.14}
\beta \langle \rho({\bf r}) \rho({\bf r}') \rangle =
\frac{1}{4\pi} \left( \frac{1}{\epsilon_0} - 1 \right)
\Delta \delta({\bf r}-{\bf r}') .
\end{equation}    
The second-moment sum rule (\ref{3.11}) follows from this expression
after an integration by parts,
\begin{eqnarray}
\int {\rm d}({\bf r}-{\bf r'})\, ({\bf r}-{\bf r}')^2
\beta \langle \rho({\bf r}) \rho({\bf r}') \rangle & = &
\frac{6}{4\pi} \left( \frac{1}{\epsilon_0} - 1 \right)
\int {\rm d}({\bf r}-{\bf r'})\, \delta({\bf r}-{\bf r}') \nonumber \\
& = &  \frac{3}{2\pi} \left( \frac{1}{\epsilon_0} - 1 \right) . \label{3.15}
\end{eqnarray}

The formula (\ref{3.10}) provides a split of the integral value 
onto its classical and purely quantum-mechanical parts.
In the quantum regime, it can be used to generate a systematic semiclassical 
expansion of the second charge-charge moment in powers of $\beta\hbar$.
In general, the r.h.s. of (\ref{3.10}) is not expressible in terms of 
elementary functions.
Perhaps the only exception is the jellium model of conductors whose 
dielectric function is adequately described, in the long-wavelength 
limit $q\to 0$, by the Drude formula with the dissipation constant 
taken as positive infinitesimal \cite{Jackson},
\begin{equation} \label{3.16}
\epsilon(\omega) = 1 - \frac{\omega_p^2}{\omega(\omega+{\rm i}\eta)} ,
\qquad \eta\to 0^+ .
\end{equation}
Inserting (\ref{3.16}) with $\eta=0$ into (\ref{3.10}) and using
the analog of the summation formula (\ref{3.8})
\begin{equation} \label{3.17}
\sum_{j=1}^{\infty} \frac{(\alpha/\pi)^2}{j^2+(\alpha/\pi)^2}
= \frac{1}{2} \left( \alpha \coth\alpha - 1 \right)
\end{equation}
for $\alpha=\beta\hbar\omega_p/2$, we obtain
\begin{equation} \label{3.18}
\frac{1}{3} \int {\rm d}^3 r\, r^2 \beta 
\langle \hat{\rho}({\bf r}) \hat{\rho}({\bf 0}) \rangle^s = 
- \frac{1}{2\pi} g(\omega_p) .
\end{equation}
This result coincides with the microscopic finding \cite{Martin85,Martin86}.
Of course, the result (\ref{3.18}) could be derived much simpler by
inserting the Drude formula for the jellium (\ref{3.16}) directly into 
(\ref{3.5}) and then calculating the ensuing integral by using 
the Weierstrass theorem, see Eq. (\ref{3.19}) which follows.
The importance of the formula (\ref{3.10}) consists in the fact that it 
provides the division of the integral onto its classical and purely 
quantum-mechanical parts for {\em all} media. 

\subsection{Time-dependent case} 
In the case of the jellium model with the dielectric function 
(\ref{3.16}), the Weierstrass theorem
\begin{equation} \label{3.19}
\lim_{\eta\to 0^+} \frac{1}{x\pm {\rm i}\eta} =
{\cal P}\left( \frac{1}{x} \right) \mp {\rm i}\pi \delta(x)
\end{equation}
(${\cal P}$ denotes the Cauchy principal value) implies
\begin{equation} \label{3.20}
{\rm Im} \frac{1}{\epsilon(\omega)} = - \frac{\pi\omega_p}{2} {\rm sgn}(\omega)
\left[ \delta(\omega-\omega_p) + \delta(\omega+\omega_p) \right] .
\end{equation}
Substituting this expression into Eq. (\ref{3.4}), we arrive at
\begin{equation} \label{3.21}
\frac{1}{3} \int {\rm d}^3 r\,  r^2 \beta 
\langle \hat{\rho}(t,{\bf r}) \hat{\rho}(0,{\bf 0}) \rangle^s
= - \frac{1}{2\pi} g(\omega_p) \cos(\omega_p t) .
\end{equation}

The time-dependent second-moment sum rule for the unsymmetrized charge-charge
density correlation function has been obtained microscopically in Refs. 
\cite{Martin88,John93},
\begin{equation} \label{3.22}
\frac{1}{3} \int {\rm d}^3 r\, r^2 \beta 
\langle \hat{\rho}(t,{\bf r}) \hat{\rho}(0,{\bf 0}) \rangle
= - \frac{1}{2\pi} \frac{\beta\hbar\omega_p}{2} \left(
\frac{{\rm e}^{{\rm i}\omega_p t}}{{\rm e}^{\beta\hbar\omega_p}-1} +
\frac{{\rm e}^{-{\rm i}\omega_p t}}{1-{\rm e}^{-\beta\hbar\omega_p}} \right) .
\end{equation}
It is easy to verify that the symmetrization of this expression
leads directly to the result (\ref{3.21}).

The general formula (\ref{3.4}) is applicable to any conducting or
dielectric medium, but since no other than jellium microscopic results 
are available we shall omit its analysis.

\renewcommand{\theequation}{4.\arabic{equation}}
\setcounter{equation}{0}

\section{Charge-current density correlations} \label{Sect.4}
As concerns the charge-current density correlation function, using 
the spectral representations (\ref{2.14}) and (\ref{2.18}) we obtain
\begin{eqnarray} \label{4.1}
\beta \langle \rho j_k \rangle_{\omega,{\bf q}}^s  & = & 
\frac{\beta\omega}{(4\pi)^2} \sum_j q_j \left[
q_k \sum_l q_l \langle A_j A_l \rangle_{\omega,{\bf q}}^s
- \left( q^2-\frac{\omega^2}{c^2} \right)
\langle A_j A_k \rangle_{\omega,{\bf q}}^s \right] \nonumber \\
& = & - q_k \frac{g(\omega)}{2\pi} {\rm Im}\, \frac{1}{\epsilon(\omega)} .
\end{eqnarray}
As a check of the validity of this relation, we first multiply its both sides 
by $q_k$, then sum over $k$ and finally use the spectral version of
the continuity equation (\ref{2.19}) to arrive at (\ref{3.1}).
 
The inverse Fourier transform of (\ref{4.1}) leads to two conditions:
\begin{equation} \label{4.2}
\int {\rm d}^3 r \beta 
\langle \hat{\rho}(t,{\bf r}) \hat{j}_k(0,{\bf 0}) \rangle^s = 0
\end{equation}
and
\begin{equation} \label{4.3}
\int {\rm d}^3 r\, x_l 
\beta \langle \hat{\rho}(t,{\bf r}) \hat{j}_k(0,{\bf 0}) \rangle^s
= \delta_{kl} \int_{-\infty}^{\infty} {\rm d}\omega\, 
\exp(-{\rm i}\omega t) \frac{g(\omega)}{4\pi^2{\rm i}} 
{\rm Im}\, \frac{1}{\epsilon(\omega)} .
\end{equation}
Note that the sum rule (\ref{4.3}) fixes the correlation function of 
the $k$th current component at time $0$ and the corresponding $l$th 
component of the total dipole moment at time $t$.  
The conditions (\ref{4.2}) and (\ref{4.3}) have the same form in both 
nonretarded and retarded regimes.

The static $t=0$ version of the sum rule (\ref{4.3}) is trivial.
Since $g(\omega) {\rm Im}\, \epsilon^{-1}(\omega)$ is an odd function
of $\omega$, we have
\begin{equation} \label{4.4}
\int {\rm d}^3 r\, x_l 
\beta \langle \hat{\rho}({\bf r}) \hat{j}_k({\bf 0}) \rangle^s = 0 .
\end{equation}

Let us now consider the time-dependent version of the condition (\ref{4.3}) 
for the jellium.
Substituting the formula (\ref{3.20}) into (\ref{4.3}), we get 
\begin{equation} \label{4.5}
\int {\rm d}^3 r\, x_l 
\beta \langle \hat{\rho}(t,{\bf r}) \hat{j}_k(0,{\bf 0}) \rangle^s
= \delta_{kl} \frac{g(\omega_p)\omega_p}{4\pi} \sin(\omega_p t) .
\end{equation}
The same result has been obtained, after the symmetrization, 
by microscopic approaches \cite{John93,Alastuey89}.

\renewcommand{\theequation}{5.\arabic{equation}}
\setcounter{equation}{0}

\section{Current-current density correlations} \label{Sect.5}
For the current-current density correlation function, using the spectral 
representation (\ref{2.18}) we obtain after some algebra that
\begin{equation} \label{5.1}
\beta \langle j_k j_l \rangle_{\omega,{\bf q}}^s =
- \frac{g(\omega)\omega}{2\pi} \left[ 
\left( \delta_{kl} - \frac{q_k q_l}{q^2} \right)
{\rm Im}\left\{ \frac{[1-(cq/\omega)^2]^2}{\epsilon(\omega)
-(cq/\omega)^2} \right\} + \frac{q_k q_l}{q^2} 
{\rm Im}\frac{1}{\epsilon(\omega)} \right] .
\end{equation}
As a check, multiplying both sides of this relation by $q_l$, summing
over $l$ and using the spectral version of the continuity equation 
(\ref{2.19}) we arrive at (\ref{4.1}).

The Fourier component (\ref{5.1}) has a well defined value at 
${\bf q}={\bf 0}$ since the two ``dangerous'' terms of type
$q_k q_l/q^2$ just cancel with one another at this point,
in both quantum and classical regimes.
As a result, the current-current density correlation function 
is integrable in space and its zeroth moment reads  
\begin{equation} \label{5.2}
\int {\rm d}^3 r\, 
\beta \langle \hat{j}_k(t,{\bf r}) \hat{j}_l(0,{\bf 0}) \rangle^s
= - \delta_{kl} \int_{-\infty}^{\infty} {\rm d}\omega\, 
\exp(-{\rm i}\omega t) \frac{g(\omega)\omega}{4\pi^2} 
{\rm Im}\, \frac{1}{\epsilon(\omega)} .
\end{equation}
The spatial integrability of the current-current density correlation function 
observed here is, specifically in the quantum regime, in contradiction 
with findings of microscopic theories for particles interacting through
Coulomb forces only \cite{Martin85,Martin88,John93}; 
a discussion about this subject will be given below in Sect. 6.

\subsection{Static case}
In the static $t=0$ case, the formula (\ref{5.2}) can be rewritten
in the form
\begin{equation} \label{5.3}
\int {\rm d}^3 r\,
\beta \langle \hat{j}_k({\bf r}) \hat{j}_l({\bf 0}) \rangle^s
= - \delta_{kl} \int_{-\infty}^{\infty} 
\frac{{\rm d}\omega}{\omega} {\rm Im}\, f(\omega) ,
\end{equation}
where
\begin{equation} \label{5.4}
f(\omega) = \frac{g(\omega)}{4\pi^2} 
\left[ \frac{\omega^2}{\epsilon(\omega)} - \omega^2 -\omega_p^2 \right] ; 
\end{equation}
the subtraction of the real sum $\omega^2+\omega_p^2$ from 
$\omega^2/\epsilon(\omega)$ has no effect due to the operation Im.
The $f$-function possesses the symmetry $f^*(\omega)=f(-\omega)$
and vanishes at $\vert\omega\vert\to\infty$ due to the asymptotic
relation (\ref{3.6}).
We can therefore apply the integration techniques together with the general 
analytic properties of dielectric functions in the complex frequency
upper half-plane presented in the second paragraph of Sect. 3.1, see
Eqs. (\ref{3.7})-(\ref{3.10}).
The final result reads
\begin{equation} \label{5.5}
\int {\rm d}^3 r\,
\beta \langle \hat{j}_k({\bf r}) \hat{j}_l({\bf 0}) \rangle^s
= \delta_{kl} \left\{ \frac{\omega_p^2}{4\pi} +
\frac{1}{2\pi} \sum_{j=1}^{\infty} \left[
\frac{\xi_j^2}{\epsilon({\rm i}\xi_j)} - \xi_j^2 + \omega_p^2 \right]
\right\} ,
\end{equation}
where $\xi_j=2\pi j/(\beta\hbar)$ are the (real) Matsubara frequencies.

In the high-temperature (classical) limit $\beta\hbar\omega_p\to 0$, each of 
the frequencies $\{ \xi_j \}_{j=1}^{\infty}$ is much larger than $\omega_p$, 
the corresponding terms in the sum on the r.h.s. of Eq. (\ref{5.5}) vanish and 
thus, for all conductor and dielectric media, it holds
\begin{equation} \label{5.6}
\int {\rm d}^3 r\, \beta \langle j_k({\bf r}) j_l({\bf 0}) \rangle_{\rm cl}
= \delta_{kl} \frac{\omega_p^2}{4\pi} .
\end{equation}
This classical result can be obtained in a simpler alternative way.
In the classical regime, the position-dependent vector potential of
the EM field can be eliminated from the momentum part of the Gibbs measure 
as well as from the definition of the current density by the Bohr-van Leeuwen 
theorem \cite{Bohr,Leeuwen}.
Consequently,
\begin{eqnarray}
\beta \langle j_k({\bf r}) j_l({\bf r}') \rangle_{\rm cl}
& = & \beta e^2 \left\langle \sum_{i=1}^N v_i^{(k)} \delta({\bf r}-{\bf r}_i)  
\sum_{j=1}^N v_j^{(l)} \delta({\bf r}'-{\bf r}_j) \right\rangle_{\rm cl}
\nonumber \\
& = & \delta_{kl} \beta e^2 n \delta({\bf r}-{\bf r}') 
\langle v^2 \rangle_{\rm cl} . \label{5.7}
\end{eqnarray}
The momentum part of the one-particle Boltzmann factor is proportional to
$\exp(-\beta m v^2/2)$, so that $\langle v^2 \rangle_{\rm cl} = 1/(\beta m)$.
With regard to the definition of the plasma frequency (\ref{3.6})
we conclude that 
\begin{equation} \label{5.8}
\beta \langle j_k({\bf r}) j_l({\bf r}') \rangle_{\rm cl}
= \delta_{kl} \frac{\omega_p^2}{4\pi} \delta({\bf r}-{\bf r}') .
\end{equation}
This result has been obtained for the classical jellium longtime ago 
by Felderhof \cite{Felderhof64,Felderhof65}.
The classical sum rule (\ref{5.6}) is a trivial consequence of
the exact relation (\ref{5.8}).
Moreover, higher moments of the current-current density correlation function
vanish in the classical limit,
\begin{equation} \label{5.9}
\int {\rm d}^3 r\, r^{2 i}
\beta \langle j_k({\bf r}) j_l({\bf 0}) \rangle_{\rm cl} = 0 ,
\qquad i = 1,2,\ldots .
\end{equation}

In the quantum regime, the formula (\ref{5.5}) can be used to generate 
a systematic semiclassical expansion of the zeroth moment of the
current-current density correlation in powers of $\beta\hbar$.
For the jellium conductor with the Drude dielectric function (\ref{3.16}),
using the summation formula (\ref{3.17}) we obtain explicitly
\begin{equation} \label{5.10}
\int {\rm d}^3 r\,
\beta \langle \hat{j}_k({\bf r}) \hat{j}_l({\bf 0}) \rangle^s
= \delta_{kl} \frac{\omega_p^2}{4\pi} g(\omega_p) .
\end{equation}
This result is not confirmed by microscopic theories for particles 
interacting only through Coulomb forces \cite{Martin85,Martin88,John93} 
which predict the spatial nonintegrability of the current-current density 
correlation function in the quantum regime (see Sect. 6).

\subsection{Time-dependent case}
In the special case of the jellium, using the relation (\ref{3.20}) 
in Eq. (\ref{5.2}), the generalization of the static formula (\ref{5.10}) 
to time difference $t$ is found to be
\begin{equation} \label{5.11}
\int {\rm d}^3 r\,
\beta \langle \hat{j}_k(t,{\bf r}) \hat{j}_l(0,{\bf 0}) \rangle^s
= \delta_{kl} \frac{\omega_p^2}{4\pi} g(\omega_p) \cos(\omega_p t) .
\end{equation}

\renewcommand{\theequation}{6.\arabic{equation}}
\setcounter{equation}{0}

\section{Problematic points in current-current 
fluctuations} \label{Sect.6}
There are certain problematic points in the results for the current-current
density correlation function, in the fluctuational theory presented here 
as well as in microscopic approaches \cite{Martin85,Martin88,John93}.

We start with an obvious inconsistency having the origin in the formula 
(\ref{5.1}). 
We shall consider the jellium with the dielectric constant (\ref{3.16}).
Using the relation (\ref{3.20}) and the similar one
\begin{equation} \label{6.1}
{\rm Im}\left\{ \frac{[1-(cq/\omega)^2]^2}{\epsilon(\omega)
-(cq/\omega)^2} \right\} 
= - \frac{\pi\omega_p^4}{2\tilde{\omega}_p^3} {\rm sgn}(\omega) \left[ 
\delta(\omega-\tilde{\omega}_p) + \delta(\omega+\tilde{\omega}_p) \right] 
\end{equation}
with a modified plasma frequency $\tilde{\omega}_p = \sqrt{\omega_p^2+(cq)^2}$,
the partial inverse Fourier transformation of (\ref{5.1}) in time leads to
\begin{equation} \label{6.2}
\beta \langle j_k j_l \rangle_{t,q}^s = 
\frac{\omega_p^2}{4\pi} \left[ \frac{\omega_p^2}{\tilde{\omega}_p^2} 
g(\tilde{\omega}_p) \cos(\tilde{\omega}_p t) 
\left( \delta_{kl} -\frac{q_k q_l}{q^2} \right) +
g(\omega_p) \cos(\omega_p t) \frac{q_k q_l}{q^2} \right] .
\end{equation}
In the static $t=0$ case and in the classical limit, which corresponds to 
setting $g(\omega_p) = g(\tilde{\omega}_p) = 1$, this formula is equivalent to
\begin{eqnarray}
\int {\rm d}^3 r\, {\rm e}^{-{\rm i}{\bf q}\cdot {\bf r}}
\beta \langle j_k({\bf r}) j_l({\bf 0}) \rangle_{\rm cl}
& = & \frac{\omega_p^2}{4\pi} \left[ \frac{\omega_p^2}{\omega_p^2+(cq)^2} 
\left( \delta_{kl} -\frac{q_k q_l}{q^2} \right) + \frac{q_k q_l}{q^2} \right] 
\nonumber \\ & = &
\delta_{kl} \frac{\omega_p^2}{4\pi} +
\frac{c^2}{4\pi} ( q_k q_l - \delta_{kl} q^2 ) + O(q^4) , \label{6.3}
\end{eqnarray}
which is an infinite series in the wave-vector components.
On the other hand, as follows from the exact analysis between 
Eqs. (\ref{5.6})-(\ref{5.9}), only the zeroth moment of the classical
current-current density correlation function is nonzero, so that it must hold
\begin{equation} \label{6.4}
\int {\rm d}^3 r\, {\rm e}^{-{\rm i}{\bf q}\cdot {\bf r}}
\beta \langle j_k({\bf r}) j_l({\bf 0}) \rangle_{\rm cl}
= \delta_{kl} \frac{\omega_p^2}{4\pi} .
\end{equation}
Our expression (\ref{6.3}) coincides with this exact formula only
in the leading absolute term which was in the center of interest in Sect. 5.
The erroneous presence of higher-order terms is intuitively related to 
the fact that Rytov's fluctuational theory works with the local dielectric 
function $\epsilon(\omega)$ which is the long-wavelength limit of 
the ${\bf q}$-dependent one $\epsilon(\omega,{\bf q})$, 
$\epsilon(\omega) = \lim_{{\bf q}\to {\bf 0}} \epsilon(\omega,{\bf q})$.
A complete theory should involve the whole function $\epsilon(\omega,{\bf q})$.
We can anticipate the validity of Rytov's fluctuational theory only in 
the leading ${\bf q}$ order, in which the ${\bf q}$-dependence of 
$\epsilon(\omega,{\bf q})$ can be ignored without any approximation.
In fact, without saying it, this strategy has already been adopted.

In microscopic approaches for purely Coulomb systems
\cite{Martin85,Martin88,John93}, the obtained result for the (partial) 
Fourier transform of the current-current density correlation function
\begin{equation} \label{6.5}
\beta \langle j_k j_l \rangle_{t,q}^s = 
\frac{\omega_p^2}{4\pi} \left[ \left( \delta_{kl} -\frac{q_k q_l}{q^2} \right) 
+ g(\omega_p) \cos(\omega_p t) \frac{q_k q_l}{q^2} \right]
\end{equation}
differs from the present one (\ref{6.2}) in the prefactor to the transverse 
component which does not depend neither on time $t$ nor on $\beta\hbar$.
The expression (\ref{6.5}) works well in the static $(t=0)$ classical 
$(\beta\hbar\to 0, g(\omega_p)\equiv 1)$ regime, where it reproduces 
the exact formula (\ref{6.4}).
As soon as the time difference $t\ne 0$ or the regime is quantum
$(\beta\hbar\ne 0, g(\omega_p)>1)$, the two ``dangerous'' terms of type 
$q_k q_l/q^2$ do not cancel with one another in (\ref{6.5}).
Since
\begin{equation} \label{6.6}
\int \frac{{\rm d}^3q}{(2\pi)^3}\, {\rm e}^{{\rm i}{\bf q}\cdot {\bf r}}
\frac{q_k q_l}{q^2} =
- \frac{\partial^2}{\partial x_k \partial x_l} 
\int \frac{{\rm d}^3q}{(2\pi)^3}\, {\rm e}^{{\rm i}{\bf q}\cdot {\bf r}}
\frac{1}{q^2} 
= - \frac{\partial^2}{\partial x_k \partial x_l} \frac{1}{4\pi r} ,
\end{equation} 
the current-current density correlation function has a nonintegrable algebraic 
decay $1/r^3$ at large distances.

The discrepancy between our (retarded) and microscopic (non-retarded) results
for the current-current density correlations is probably analogous to that 
for the induced electric potential and field correlation functions
\cite{Jancovici06,LP}.
Our analysis does not exclude an algebraic long-range behavior of the
current-current density correlation function, but this decay must be fast
enough to ensure the spatial integrability of the correlation function. 

\renewcommand{\theequation}{7.\arabic{equation}}
\setcounter{equation}{0}

\section{Classical limit of current-current fluctuations} \label{Sect.7}
The classical limit of the current-current density correlation function 
for the jellium was investigated longtime ago by Felderhof, 
in the nonretarded regime \cite{Felderhof64} as well as for the retarded 
plasma fully coupled to the classical radiation \cite{Felderhof65}.
Changing the notation ${\bf k}\to {\bf q}$ and using the substitution
$s=m\omega/q$, his final formula (13.5) in  Ref. \cite{Felderhof65} can be 
rewritten after some straightforward algebra as follows 
\begin{equation} \label{7.1}
\beta \langle j_k j_l \rangle_{\omega,{\bf q}} =
- \frac{\omega}{2\pi} \left[ \left( \delta_{kl} - \frac{q_k q_l}{q^2} \right)
{\rm Im}\left\{ \frac{[1-(cq/\omega)^2]^2}{Z_1 q^2/(m\omega)^2}
\right\} + \frac{q_k q_l}{q^2} {\rm Im}\frac{1}{Z_0} \right] ,
\end{equation}
where
\begin{equation} \label{7.2}
Z_0(\omega,q) = 1 - \omega_p^2 \int_{-\infty}^{\infty} {\rm d}x\,
\frac{1}{(\omega+{\rm i}\eta/2-x)^2} \delta_a(x) 
\end{equation}
and
\begin{equation} \label{7.3}
Z_1(\omega,q) = \left( \frac{m\omega}{q} \right)^2 \left[
1 - \left( \frac{cq}{\omega} \right)^2 -
\frac{\omega_p^2}{\omega} \int_{-\infty}^{\infty} {\rm d}x\,
\frac{1}{\omega+{\rm i}\eta-x} \delta_a(x) \right]
\end{equation}
with $\eta\to 0^+$;
the function $\delta_a(x)$ and the parameter $a$ are defined by
\begin{equation} \label{7.4}
\delta_a(x) = \frac{1}{\sqrt{\pi}a} {\rm e}^{-x^2/a^2} , \qquad
a = q \sqrt{\frac{2}{\beta m}} .
\end{equation}

In the limit ${\bf q}\to {\bf 0}$ ($a\to 0$), $\delta_a(x)$
becomes one of the standard definitions of the $\delta$-function.
Consequently,
\begin{equation} \label{7.5}
Z_0(\omega,q) \mathop{\sim}_{q\to 0} \epsilon(\omega) , \qquad
Z_1(\omega,q) \left( \frac{q}{m\omega} \right)^2 \mathop{\sim}_{q\to 0} 
\epsilon(\omega) - \left( \frac{cq}{\omega} \right)^2 ,
\end{equation}
where $\epsilon(\omega)$ is the dielectric function of 
the jellium (\ref{3.16}).
Inserting these expressions into (\ref{7.1}) and comparing the resulting 
formula with the classical version ($g(\omega)\equiv 1$) of 
the current-current density correlation function (\ref{5.1}), we see the
equivalence of the two approaches in the considered limit 
${\bf q}\to {\bf 0}$.  
However, when ${\bf q}$ is finite, the approaches give different
results.
Felderhof's classical formula turns out to be correct for an arbitrary
value of ${\bf q}$ since it leads to the exact relation (\ref{5.8}) or
its equivalent (\ref{6.4}) \cite{Felderhof65}.
From the discussion in Sect. 6 we know that this is no longer true
for the classical version of our formula (\ref{5.1}).
This fact might be a motivation for searching new improvements of Rytov's
fluctuational theory which reproduce classical results for any value
of $q$ and simultaneously describe adequately correlation functions
in the quantum regime.

\renewcommand{\theequation}{8.\arabic{equation}}
\setcounter{equation}{0}

\section{Conclusion} \label{Sect.8}
In this paper, by using Rytov's fluctuational theory we have derived 
a series of known or new sum rules for the bulk correlation functions
of charge and current densities in conductor and dielectric media, 
fully coupled to the radiation field (retarded regime).  
Few important facts are worth to mention.

In the Fourier ${\bf q}$ space, we can anticipate the validity of Rytov's 
macroscopic approach only in the leading ${\bf q}$ order. 
In this order, the ${\bf q}$-dependence of the dielectric function
$\epsilon(\omega,{\bf q})$ can be ignored without any approximation.

In the case of the static $t=0$ sum rules, an analysis in the complex
frequency upper half-plane was used to split these sum rules onto their 
classical and purely quantum-mechanical parts.
The classical part coincides, for both conductors and dielectrics, 
with the results obtained previously by other methods.
The quantum-mechanical part was checked on the jellium model of conductor.

As concerns the time-dependent sum rules, their validity was tested
on the jellium, too.
The retarded results for the charge-charge and charge-current density 
correlation functions coincide with the non-retarded ones obtained 
previously by microscopic approaches to the jellium model.
The general formulas are applicable to any media, but no other than
jellium microscopic results are available to make a comparison.

The most interesting quantity is the current-current density 
correlation function.
Within microscopic theories in the non-retarded regime, this correlation 
function turns out to be spatially integrable in the classical limit, 
but it becomes nonintegrable in the quantum case due to an algebraic decay 
$1/r^3$ at large distances.
This contradicts our retarded prediction (\ref{5.2}) about a finite value of 
the integral over the space, in both classical and quantum cases.

A generalization of the present results to inhomogeneous 
situations, like a surface contact between two distinct media, 
is left for the future.

\begin{acknowledgements}
It is a pleasure to thank Bernard Jancovici for fruitful discussions
on the subject of this work and careful reading of the manuscript. 
I am grateful to LPT for very kind invitation and hospitality.
The support received from the European Science Foundation 
(``Methods of Integrable Systems, Geometry, Applied Mathematics''),
Grant VEGA No. 2/0113/2009 and CE-SAS QUTE is acknowledged. 
\end{acknowledgements}

\end{document}